\documentclass[11pt,a4paper]{article}
\usepackage{amsfonts}
\usepackage{graphicx}
\usepackage{amsmath}
\usepackage{hyperref}
\usepackage{enumerate}

\makeatletter \@addtoreset{equation}{section}

\newcommand{\be}{\begin{equation}}
\newcommand{\ee}{\end{equation}}
\newcommand{\bea}{\begin{eqnarray}}
\newcommand{\eea}{\end{eqnarray}}
\begin{document}

\title{{ \begin{flushright}
{\normalsize\small LPHE-MS-11-07/ CPM-11-07/ FPAUO-11/07}
\end{flushright} }\bf On   Chern-Simons Quivers  and Toric Geometry}
\author{ Adil  Belhaj$^{1,2,6}$\thanks{belhaj@unizar.es},  Pablo Diaz$^{3}$\thanks{pablo.diazbenito@wits.ac.za},
 Maria Pilar  Garcia del Moral$^{4}$\thanks{garciamormaria@uniovi.es},
 Antonio  Segui$^{5}$\thanks{segui@unizar.es}\hspace*{-15pt} \\
%EndAName
\\
{\small $^{1}$Lab Phys Hautes Energies, Modelisation et Simulation, Facult%
\'{e} des Sciences, Rabat, Morocco} \\
{\small $^{2}$Centre of Physics and Mathematics, CPM-CNESTEN, Rabat, Morocco } \\
{\small $^{3}$ National Institute for Theoretical Physics, University of  Witwatersrand, South Africa } \\
{\small $^{4}$Departamento de F\'isica, Universidad de Oviedo, Avda Calvo Sotelo
18. 33007 Oviedo, Spain } \\
{\small $^{5}$Departamento de F\'isica Te\'orica, Universidad de
Zaragoza,
E-50009-Zaragoza, Spain}\\
{\small $^{6}$Groupement National de Physique des Hautes Energies,
Si\`{e}ge focal: FSR, Rabat, Morocco }} \maketitle

\begin{abstract}
We discuss  a class of 3-dimensional $\mathcal{N}=4$  Chern-Simons
(CS) quiver gauge models obtained    from  M-theory
compactifications
 on singular  complex 4-dimensional hyper-K\"{a}hler (HK) manifolds, which are
   realized explicitly as a cotangent bundle over   two-Fano  toric varieties $V^2$. The corresponding
CS gauge models are encoded in  quivers   similar to  toric diagrams
of   $V^2$.  Using  toric geometry, it is  shown that the
constraints on  CS levels  can be related to   toric equations
determining  $V^2$.

\end{abstract}

%\newpage
%\tableofcontents \thispagestyle{empty} \newpage \setcounter{page}{1}
\newpage

\section{Introduction}
In string theory,  the properties of gauge quiver  sectors localized
on worldvolume  of D-branes at
 singular points of Calabi-Yau  manifolds  in the low energy theory are determined by the local geometry
 around those singularities \cite{HI,adil}.  In the $\mathcal{N}=1$  supersymmetric case for instance,
    the field content of D-brane worldvolumes
 can be encoded in a quiver diagram, where the $i^{th}$ node represents the $\mbox{U}(N_i)$ factor of the
  gauge group and there are  oriented arrows from the $i^{th}$ to the $j^{th}$ node corresponding
  to $\mathcal{N}=1$ chiral multiplets in the $(N_i,\overline{N}_j)$ representation \cite{angel}.
  The matter
  content  is given by bi-fundamental hypermultiplets
in the representations $(N_i,\overline{N}_j)$ of the gauge group.
The superpotential is obtained by
 restricting the chiral multiplets to the invariant ones. The
   moduli spaces  of the vacua associated  with  such quivers that corresponds to the space of
   possible locations of the D$p$-brane, are  usually  isomorphic to orbifold geometries.
   In connection with AdS/CFT correspondence in four dimensions,  the  dual models  are given by
     spaces of the form $AdS_5\times M_5$, where
$M_5=S^5/\Gamma$  with $\Gamma$ being a discrete subgroup of SO(6)
\cite{KaS}. In particular,  these can be obtained by placing
D3-branes  at ADE singularities of Calabi-Yau threefolds
generalizing the conifold singularity \cite{GNS}.

 Recently,
3-dimensional Chern-Simons (CS) quivers have  attracted much
attention and have been investigated from various points of view in
type II superstrings and M-theory
 compactifications \cite{MS,HLLP,HZ,GW}.  In particular,  it has been pointed out that 3-dimensional
  $\mathcal{N}=6$ CS  quiver  with  $\mbox{U}(N)_k \times  \mbox{U}(N)_{-k}$  gauge symmetry is dual to  M-theory
  propagating on $AdS_4\times S^7/Z_k$, with an appropriate amount of fluxes, or to type IIA superstring on
    $AdS_4\times \bf CP^3$ for large number of $k,N$ with $k\ge N$ corresponding to the weakly interacting regime
     \cite{ABJM}. In the decoupling limit, the corresponding  CFT$_3$  is generated by
the action of multiple M2-branes placed at the orbifold $ C^4/Z_k$.
This result  can be considered a nice example for understanding the
AdS/CFT correspondence in  three dimensions.

This   analysis has been extended in several ways and  applied for a
large class of examples of CS quivers which are proposed to be dual
theories to theories on non trivial  seven dimensional   manifolds.
In particular,  it has been suggested that  some $\mathcal{N}=2$
models   with  M2's    are  dual to seven dimensional toric
Sasaki-Einstein manifolds Y$_7$,
 which are considered as the base of  a Calabi-Yau cone $C($Y$_7)$. The corresponding  Chern-Simons-Matter
 (CSM)  theories  are quivers with different number of supersymmetries and their  classical Vacuum Moduli Spaces
  (VMS) have been analyzed in several works, take for example   the case  of
  \cite{MS}. In particular, it has also been shown  how to extract toric data of the Calabi-Yau 4-fold $C$(Y$_7)$ to  describe
  the corresponding quiver theories. These $\mathcal{N}=2$ Chern-Simons share many similarities  with four-dimensional
  quiver gauge models preserving only four supercharges, obtained  from type II superstrings on
Calabi-Yau threefolds  or M-theory on G2 manifolds,  although their
gauge symmetries and D-term conditions are modified \cite{diego,
HI}. In \cite{schwarz},  a realization of Maldacena conjecture
\cite{maldacena} in three dimensions (AdS$_4$/CFT$_3$)
 was searched for. There,  it was there pointed out   the necessity  of  having  $\mathcal{N}\ge 4$ superconformal
  CS  theories to capture the low energy description of the infrared fixed point of the worldvolume gauge theory
  describing multiple M2-branes. The $\mathcal{N}=8$ realization, corresponds to the description of multiple M2-branes
   in flat space \cite{rey}. In the strongly coupled regime the associated  physics  corresponds to the worldvolume description of
    multiple M2-branes\footnote{For a High Energy description (opposite regime to the decoupling limit) of multiple M2-branes
    with an arbitrary number of colors and without imposing confomality see \cite{gmr}.} in distinction with type  IIB  superstring case
     that describes the weakly coupled regime.  When the 3d theory is dual to M-theory on
     $AdS_4\times Y_7$ then  the Vacuum
      Moduli Space (VMS) of  the 3d theory $\mathcal{M}_{3d}$ should coincide with the CY 4-fold cone $C(Y_7)$,
       consisting  of
a quiver diagram associated  with  the D-terms, a superpotential and
CS levels ($k_i$).

Gaiotto-Witten  (GW) \cite{GW}  and \cite{HLLP} exhausted all the 3d
CSM quiver compatible $\mathcal{N}=4$
 conformal supersymmetries. For other formulations developped as twisted examples of these two types of constructions see
 \cite{imamura, KS}. In the case of $\mathcal{N}=4$ SCM theories the Vacuum Moduli space  (VMS) also receives quantum corrections \cite{imamura,imamura2}
 that we do not consider here.

The aim of this work  is to contribute to these activities by
considering a       class of 3-dimensional
 CSM quiver  gauge theories from geometric
data of  a particular  M-theory compactification. In this way, the
gauge group and matter content of
 the resulting
models are obtained from the singularities of complex 4-dimensional
hyper-K\"{a}hler (HK) manifold. The geometry is realized explicitly
as a cotangent bundle over  complex  two-dimensional toric
varieties.
 It enables one to represent the corresponding
CS gauge models by quiver diagrams similar to  toric graphs of
two-Fano  toric varieties $V^2$. One considers the quiver  to be
composed of a set of vertices, each of which is associated with
$\mbox{U(N)}$ gauge group factor, and for each pair of   vertices
with matter in bi-fundamental representations, where the  vertices
are connected by  lines required by toric geometry.  This diagram
encodes some  information on the world-volume  of M2-branes probing
the above  toric Calabi-Yau four-fold (hyper) cones. As an
illustration, we consider triangle and  rectangular  quiver gauge
models by introducing projectives spaces,
 Hirzebrouch surfaces and  del Pezzo surfaces  in the base of the cotangent fibration in M-theory backgrounds.
 Using  toric geometry,   we show
that the CS constraints can be converted into  toric equations
defining  two-Fano  toric varieties. We have taken as a departing
point, the D-term constrains of the  conformal SCM (Super
Chern-Simons Matter) theory to construct a quiver and together with
the superpotential (F-term constrains), to obtain a toric
Hyperkahler $\mathcal{N}=4$ diagram which has only two directions
compact. We compare these constructions with the GW models, by
imposing restrictions on the allowed values for the CSM levels.
%%%%%%%%%%%%%%%%%%%%%%%%%%%%%%%%%%%%%%%%%%%%%%%%%%%%%%%%%%%%%%%%%%%%%%%%%%%%%%%%%%%%%%
%%%%%%%%%%%%%%%%%%%%%%%%%%%%%%%%%%%%%%%%%%%%%%%%%%%%%%%%%%%%%%%%%%%%%%%%%%%%%%%%%%%%%%%%
\section{Quiver Chern-Simons Theories}
In \cite{BL,G},  a $\mathcal{N}=8$ M2-branes lagrangian was
constructing  trying to describe the  IR limit of a 3-dimensional
CFT (CFT$_3$) to realize
 the Maldacena conjecture  in three dimensions (AdS$_4$/CFT$_3$) but restricted just to $\mathcal{N}=2$ branes. However, the difficulty of this correspondence  is due the lack of understanding
  of a IR fixed point of a conformal  field theory of multiple  of M2-branes, in particular probing  Calabi-Yau  fourfolds.
   It was pointed out that it was needed a supersymmetry  $\mathcal{N}\geq 2$  to realize such a limit.
   The authors of
    \cite{ABJM},  constructed  a $\mathcal{N}=6$ Chern-Simons with  $\mbox{U}(N)\times \mbox{U}(N)$
     as the worldvolume of $N$ M2-branes placed at the $C^4/Z_k$ orbifold.  As a quiver
diagram, this model,  which is known by
      the ABJM theory,  consists of two nodes. To
each node we associate  a  gauge factor $\mbox{U}(N)$. These have  a
Chern-Simons lagrangian with levels $k$ and $-k$. The
$\mathcal{N}=4$ Superconformal Chern Simons theories was constructed
by \cite{GW} with $SO(4)$ R-symmetry and
 $OSp(4\vert 4)$ conformal symmetry.  The associated target space corresponds to a noncompact toric  hyperkahler manifold.
 It corresponds to the description of multiple M2-branes once that the Fundamental Identity\footnote {The \emph{FI} condition
  is the generalization of Jacobi indentity to Fillipov algebras.} is imposed to guarantee invariance under the superconformal
  Lie algebra. This Fundamental Identity imposes constrains to the Chern Simons level of the gauge groups to appear in alternating
  pairs $(k,-k)$. In such a way that the unique nonabelian gauge groups for a single type of hypermultiplets, were proved to
  be: $\mbox{U}(N)_{k}\times \mbox{U}(M)_{-k}$ and $\mbox{O}(N)_{k}\times \mbox{Sp}(M)_{-k}$ or direct sums in blocks of them.
  Matter $(q_{\alpha}�,\Psi_{\dot{\alpha}} ^A)$ are in hypermultiplets in the bifundamental. The associated quiver is linear
  in such a way that  for gauge groups $\mbox{U}(N_1)_{k_1}\times \mbox{U}(N_2)_{k_2}\times \mbox{U}(N_3)_{k_3}$
   the levels are imposed to be $k_1=k,k_2=0, k_3= -k$. The associated toric manifolds  has been  shown to be:
\bea \mbox{U}(N_1)\times \mbox{U}(N_2)\times
\mbox{U}(N_3)/\mbox{U}(N_2) \eea In general for an arbitrary number
of gauge groups the associated quiver diagram has $k_1=k,
k_2=k_3=\dots=k_{s-1}=0, k_s=-k$. In \cite{HLLP},  the GW theory was
generalized by adding a twisted hypermultiplet to the GW
construction in order to include links among the different pairs of
nodes. The two types of hypermultiplets alternate among the gauge
groups. The resulting quiver can be linear or circular with multiple
nodes. The simplest case corresponds to the BLG model with a
modified R-symmetry $SO(4)$. When the gauge groups are
$\mbox{U}(N_i)$, the quivers are considered as   the Dynkin diagrams
of  $A_n$ series and $K^{mn}=(-1)^I k$ with $I\in Z$.  The authors
of \cite{GW} and \cite{HLLP} classify all the toric manifolds for a
GLSM with $\mathcal{N}=4$ representing the low energy description of
multiple M2 branes on the hyperkahler construction.

In \cite{imamura},  it is analyzed a case of a $\mathcal{N}=4$ CS
theory with auxiliary vector multiplets. It has been  considered
quiral fields in the bifundamental  and the Chern Simons coupling
generalized to
 be $k_I=\frac{k}{2}(s_I-s_{I-1})$ with $s_I=\pm 1$ $k>0$ allowing then some of the levels to vanish, $k_I=0$.
 This breaks the superconformal Lie algebra and the dual theory is no longer associated  with  M2-branes but with type IIB  superstring in the presence
  of  $N$ D3-branes intersecting on a circle $S^1$ and $n$ 5-branes intersecting the D3 along the $S^1$.
The nonlinear sigma models can be constructed as GW model  indicated
by performing the hyperkahler quotient
 of all nodes of the corresponding linear quiver diagram except  for those gauge groups of the extreme. In \cite{KS},  it has been given
  a recipe to construct the general nonlinear sigma models with compact gauge group and whose target space is
  the contangent fibration of a flag manifold $T^* (CP^1)$ for hyperkahler $\mathcal{N}=4$ superconformal CSM theories in $d=3$. \newline

There have appeared many models extending    ABJM,   and  describing
three dimensional  $\mathcal{N}=2$ quiver Chern-Simon theories. They
are conjectured  to be    gauge field   duals    of  $AdS_4$
background  in type IIA   superstring and
   M-theory compactifications.  These   CS quiver
theories can  be  interpreted  in terms of M2-branes placed at toric
four-folds  singularities. The simple model is described by an
abelian gauge symmetry $\mbox{U}(1)_{k_1}\times
\mbox{U}(1)_{k_2}\times \ldots \times \mbox{U}(1)_{k_n}$,  where $
k_i$ denote Chern-Simons levels  for each factor $\mbox{U}(1)$.
These models can be encoded in  a graph  formed by $n$ vertices
where each  gauge group factor $\mbox{U}(1)$ is  associated  with a
vertex   while the matter is represented by the link  between
vertices.   Using similarity with $\mathcal{N}=1$ quivers
 in four dimensions, the corresponding toric  moduli space  have been discussed  in  \cite{MS}.  Among others,
  it  has  been shown that  Chern-Simons levels $ k_i$ are  constrained by
\begin{equation}
\label{con} \sum_ik_i=0.
\end{equation}
This is a necessary condition for the moduli space to be a four
complex dimensional. For the general case where the gauge group  is
$ \prod_i\mbox{U}(N_i)_{k_i}$,  the constraints on  $k_i$  read as
\begin{equation}
\label{cong} \sum_ik_iN_i=0.
\end{equation}
In what follows, we will show that  these constraints can be solved
using toric geometry equations.
%%%%%%%%%%%%%%%%%%%%%%%%%%%%%%%%%%%%%%%%%%%%%%%%%%%%%%%%%%%%%%%%%%%%%%%%%%%%%%%%%%%%%%%%%%%%
%%%%%%%%%%%%%%%%%%%%%%%%%%%%%%%%%%%%%%%%%%%%%%%%%%%%%%%%%%%%%%%%%%%%%%%%%%%%%%%%%%%%%%%%%%%%%%
\section{ Toric description  of  Chern-Simons quivers }
In this section we give a  toric description of  a class of CS
quivers obtained from  M-theory on hyperkahler backgrounds.
 In particular, we will show that
the    CS conditions  given in  eq.(\ref{con}) and eq.(\ref{cong})
can be  translated  into  nice toric algebraic equations
\cite{F,LV}. The latters appear in the standard construction of
complex manifolds using toric geometry. Recall that, a
$n$-dimensional toric variety $\cal \bf  V^n$  can
 be  represented by a  toric
     diagram (polytope) $ \Delta({{\cal \bf V}^n})$ spanned by  $ k=n+r$
     vertices $ v_i$ of an  $\bf Z^n$ lattice satisfying
      \begin{equation}
    \sum \limits _{i=1}^{n+r} q_i^a v_ i=0,\quad a=1,\ldots,r,
  \end{equation}
where $q_{i}^{a}$ are integers. For each $a$ they form the so-called
Mori vectors.  The simplest example in toric geometry, which turns
out to
   play a crucial role in  the  building
   blocks  of higher-dimensional toric
varieties,  is
 $\bf CP^1$. This geometry  has an
  $\mbox{U}(1)$ toric action ($
 z\to e^{i \theta}z$) with  two fixed points
  $v_1$ and $v_2$ on the real line. These two   points  satisfy  the following condition
\begin{equation}\label{toriccp1}
v_1+v_2=0,
\end{equation} and  they  describe
  respectively the north and south poles of $\bf CP^1$. The corresponding  polytope is just
  the segment $[v_1,v_2]$ joining the two points $v_1$ and $v_2$.
  Thus, $\bf CP^1$ can be viewed as a segment $[ v_1,v_2]$  with a
  circle on top,  where the circle vanishes at the end points $v_1$
  and $v_2$. For higher dimensional geometries, the toric  descriptions  are slightly more complicated.
   For more details see \cite{F,LV}.

Roughly speaking,  we will  be interested  in CS quivers associated
with gauge theories  obtained from M-theory on  the cotangent bundle
over two dimensional toric variety  ${\cal \bf V^2}$.
 A nice way  to describe  such M-theory backgrounds  is to
use the so-called hyper-K\"{a}hler quotient  studied in
\cite{GVW,BS1} engineered by considering  a two-dimensional
U(1)$^{r}$ sigma model with eight supercharges ($\mathcal{N}=4$) and
$r+2$ hypermultiplets. There is a SU($r+2$) global symmetry under
which
 the hypermultiplets transform
in the fundamental  representation $r+2$. This background   is
defined
  by  the following D-flatness condition
\begin{equation}
\sum_{i=1}^{r+2}q_{i}^{a}[\phi _{i}^{\alpha }{\bar{\phi}}_{i\beta
}+\phi _{i}^{\beta }{\bar{\phi}}_{i\alpha
}]=\vec{\xi}_{a}\vec{\sigma}_{\beta }^{\alpha },\;\; a=1,\ldots,r
\label{sigma4}
\end{equation}
where $q_{i}^{a}$ is a matrix charge specified later on.  $\phi
_{i}^{\alpha }$'s ($ \alpha=1,2)$ denote the  component field
doublets of each hypermultiplets ($ i=1\ldots,r+2)$. $\vec{\xi}_{a}$
are  the   Fayet-Illiopoulos (FI)  3-vector couplings rotated by the
SU(2) symmetry, and $\vec{\sigma}_{\beta }^{\alpha }$ are the traceless $%
2\times 2$ Pauli matrices.  Performing   SU(2) R-symmetry
transformations $\phi ^{\alpha }=\varepsilon ^{\alpha \beta }\phi
_{\beta },\;\overline{\phi ^{\alpha }}=\overline{\phi }_{\alpha
},\;\varepsilon _{12}=\varepsilon ^{21}=1$ and replacing  the Pauli
matrices by their expressions, the identities (\ref{sigma4}) can be
split as follows\begin{equation} \label{2.2}
\begin{array}{lcr}
\sum\limits_{i=1}^{r+2}q_{i}^{a}(|\phi _{i}^{1}|^{2}-|\phi
_{i}^{2}|^{2})
&=\xi _{a}^{3}    \\
\sum\limits_{i=1}^{r+2}q_{i}^{a}\phi _{i}^{1}\overline{\phi
}_{i}^{2} &=\xi
_{a}^{1}+i{\xi ^{2}}_{a} \\
\sum\limits_{i=1}^{r+2}q_{i}^{a}\phi _{i}^{2}\overline{\phi
}_{i}^{1} &=\xi _{a}^{1}-i{\xi ^{2}}_{a}.
\end{array}
\end{equation}
Using the fact that the
  resulting  space of  (\ref{2.2})  is invariant under   U(1)$^r$ gauge transformations,
  we get precisely an
eight-dimensional toric HK manifold.   However,
 explicit  solutions of these geometries  depend on the values of the FI
   couplings.  Taking   $\xi^1_a=\xi^2_a=0$  and
  $\xi^3_a >0$,  (\ref{2.2}) describe the
  cotangent bundle over  complex  two-dimensional toric varieties.
Indeed, if we set all $\phi^2_i=0$,   we get    two-dimensional
nonsingular toric variety $\cal \bf V^2$  defined by
\begin{equation}
\label{sigman2}
  \sum\limits_{i=1}^{2+r} q_i^a|\phi^1_i|^2 = \xi^3 _a, \quad
a=1,\ldots,r.
\end{equation}
To   connect this equation with   toric geometry,  we associate to
each field $\phi^1_i$ a vector $v_i =(v_i^1,v_i^2)$ in the standard
lattice $\bf Z^2$, such that the $v_i$ fulfill the following
relations
 \begin{equation}
\label{toricequation}
    \sum \limits _{i=1}^{2+r} q_i^a v_ i=0,\quad a=1,\ldots,r.
  \end{equation}
Up to eq.(\ref{toricequation}),  these $ k=2+r$ vertices  $ v_i$
  represent  the  toric
     diagram $ \Delta$ of ${{\cal \bf V}^2}$ ($ \Delta({{\cal \bf V}^2})$). The last equations of (\ref{2.2})  mean
  that the $\phi^2_i$'s define the cotangent   fiber
  directions over ${\cal \bf  V^2}$.  To see that, let us illustrate the idea by  using the   leading example:
 $T^*( \bf CP^2$)  contangent bundle. Indeed,
we consider $2d $ $ N =4$ supersymmetric U(1) gauge theory with one
isotriplet FI coupling $\vec{\xi}=(\xi^1,\xi^2,\xi^3)$ and  only
three hypermultiplets of charges $q^i_a=q^i=1;  i=1,2,3$.  The zero
energy states of this gauge model are obtained by solving
\begin{equation}
\label{5}
\begin{array}{lcr}
\sum\limits_{i=1}^3( |\phi^1_i|^2-|\phi^2_i|^2) &= \xi^3 \qquad &
\\
\sum\limits_{i=1}^3 \phi^1_i
\overline{\phi}_{i2}&=\xi^1+i{\xi^2}\qquad &
\\
\sum\limits_{i=1}^3 \phi^2_i
\overline{\phi}_{i1}&=\xi^1-i{\xi^2}.\qquad &
\end{array}
\end{equation}
As we have seen before,   the solutions of eqs.(\ref{5})  depend on
the values of the FI couplings. For the case where
$\xi^1=\xi^2=\xi^3=0$, the moduli space has an
$\mbox{SU}(3)\times\mbox{SU}(2)_R$ symmetry; it is a cone over a
seven manifold described by
\begin{equation}
\label{7} \sum_{i=1}^3(\varphi_{\alpha i}\overline{\varphi}_i^\beta-
\varphi_i^\beta\overline{\varphi}_{\alpha i})= \delta{ _\alpha
^\beta}.
\end{equation}
 For the case $\vec{\xi}\neq\vec{0}$, the abovementioned
$\mbox{SU}(3)\times\mbox{SU}(2)_R$ symmetry is explicitly broken
down to $\mbox{SU}(3)\times\mbox{U}(1)_R$. In the remarkable case
where  $\xi^1=\xi^2=0$  and $\xi^3$ positive definite, it is not
difficult to see that (\ref{5}) describe the cotangent bundle of
$\bf CP^2$. Putting now  ${\phi}_{i}^2=0$,  one gets
\begin{equation}
 |\phi^1_1|^2+|\phi^1_2|^2+|\phi^1_3|^2 = \xi^3
\end{equation}
defining    now the
 $\bf CP^2$  projective space. On the other hand,  with $\xi^1=\xi^2=0$
 conditions, the two last equations  of (\ref{5}) may be interpreted to mean that
$ \overline{\phi}_{2i}$ lies in the cotangent space to  $\bf CP^2$.
This can be viewed as  an extension of the canonical complex cone
over $\bf CP^2$ used in the study of $\mathcal{N}=1$  quivers
embedded  in type II superstrings \cite{HI}.

Having  constructed  eight dimensional toric HK manifolds,   the
following   will be concerned with M-theory on such manifolds. We
will try to  give  a  toric geometry   description  of  the
corresponding CS quivers. In particular, one can interpret the
constraint equations  eq.(\ref{con}) and   eq.(\ref{cong})    as
toric geometry equations describing   two dimensional toric
manifolds $V^2$. The analysis we will be using here is based on the
similarity between the  four dimensional  quivers   and CS  ones. To
indicate the ideas,  we  will focus on  CS quiver theories
associated  with sigma models on the cotangent bundle over $V^2$.
Our  toric geometry interpretation of  this  class of models relies
on  the fact that the corresponding quivers  are identified with the
toric diagram of $V^2$.  In this way, the  gauge group and
Chern-Simons levels should be related to toric data $(q_i^a, v_i)$
of $V^2$. Indeed,  a close examination of the quiver method,
 used in string theory,  reveals that there are  many similarity between such CS quivers and four dimensional  $\mathcal{N}=1$
 quivers  embedded in type II superstrings  on the line bundle of  $V^2$ using the so called $(p,q)$ brane webs
 \cite{HI,adil}.  Motivated by \cite{MS} and  based on these observations,  we give a toric interpretation of   M-theory
  CS quivers. In particular, we will interpret   the toric   data   defining $V^2$ as  constraints on CS  levels.
  In  fact precisely, the parameters $k_i$ and $N_i$   given in  eq.(\ref{con})  and eq.(\ref{cong})
   will be identified with  the toric   data  $(q_i^a, v_i)$. To see that, let us discuss a simple example where $V^2$ is
     $\bf CP^2$  defined by eq.(\ref{toricequation}) for $r=1$ and the vector charge $q_i=(1,1,1)$.  We will see  that other examples
     may be dealt with in a similar way.  The  corresponding toric    quiver (diagram) has  3  vertices, where each
      vertex  is associated with an  U($N$) gauge  symmetry factor. For this CS quiver,
      we have a level  vector $(k_1,k_2,k_3)$ such that
 \begin{equation}
\label{concp2} k_1+k_2+k_3=0.
\end{equation}
We will  show that this  constraint can be solved by exploring  the
$\bf CP^2$  toric data. Indeed, we start by  recalling  that    $\bf
CP^2$  is a complex two dimensional manifold with an $\mbox{U(1)}^2$
toric action exhibiting three fixed points $v_1$, $v_2$ and $v_3$.
Its
 polytope  $\Delta_2$ is a finite sublattice of the $Z^2$ square lattice. This polytope is  described by  the
intersection of three  $\bf CP^1$ curves defining a triangle
$(v_1v_2v_3)$ in the  $\bf R^2$ plane. Toric geometry
 requires that a  convenient
choice of the data of the three vertices  is constrained by
 \begin{equation}
\label{toriccp2} v_1+v_2+v_3=0.
\end{equation}
This equation looks  like the constraint on CS levels given in
eq.(\ref{concp2}). In fact,  eq.(\ref{concp2}) can be solved by
\begin{equation}
k_i= (v^1_i+v^2_i)k
\end{equation}
where  $k$ is  an arbitrary integer.  It is worth to give some
remarks regarding  this solution. First, the same
 result should be  also  obtained  for  the dual polytope ${\Delta_2}^\star$.  For example,  the  dual polytope  of
   ${\Delta_2}$ with vertices $(-1,-1)$, $(-1, 2)$ and $(2,-1)$
 corresponding to  $\bf CP^2$  with a degree 3 bundle on it, is
${\Delta_2}^\star$  being  bounded by $(1, 1)$, $(-1, 0)$ and
$(0,-1)$.  Since the dual polytope is useful for
 constructing mirror toric varieties, it should be interesting to make contact with the mirror maps in CS theories  discussed in
  \cite{CKS}.  Second,  it is straightforward to recover the result   concerning  $Y^{p,k}$ metrics  studied in \cite{MS}.
  For $k=0$, the connection  can be  ensured by  the solution
\begin{equation}
k_i= p({v^\star}^1_i+{v^\star}^2_i)
\end{equation}
where now ${v^\star}_i$ are  the toric vertices  of the dual
polytope of $\bf CP^2$. It worth to note that, in the  case in which
the Chern Simons levels   are associated with  a linear quiver  with
$k_1=k, k_2=0, k_3=-k$  would correspond to the GW model \cite{GW}.
\\
As we  have seen, the vector  $(k_1,k_2,k_3)$ in triangle CS quiver
is related to the  toric  vertices $(v_1,v_2,v_3)$. It is natural to
ask many  questions. One of them is as follows.  Is there any
interpretation for the  ranks $N_i$?. In what follows, we speculate
on it. For this reason, it may be useful to introduce toric
geometries with $q_i\neq 1$. In fact, we will replace   $\bf CP^2$
by  a weighted  projective spaces  ${\bf WCP^2}_{q_1, q_2,q_3}$.
 In this case, eq.(\ref{toriccp2}) becomes
\begin{equation}
\label{toricwcp2} q_1v_1+q_2v_2+q_3v_3=0.
\end{equation}
Keeping the same  analysis of   $\bf CP^2$,  eq.(\ref{cong}) can be
solved by
\begin{eqnarray}
k_i&=& (v^1_i+v^2_i)k\nonumber\\
N_i&=&q_in
\end{eqnarray}
where $n$ is arbitrary. The corresponding gauge symmetry is
\begin{equation}
\mbox{U}(q_1n)\times \mbox{U}(q_2n)\times \mbox{U}(q_3n)
\end{equation}
 This model is encoded in a  quiver with   3 vertices. Each vertex is associated with  $\mbox{U}(q_in)$ gauge group
  and  the links  are associated with $\Pi f_i$
 bifundamental  fields.  This can  be    represented by
\vspace{0.5cm}
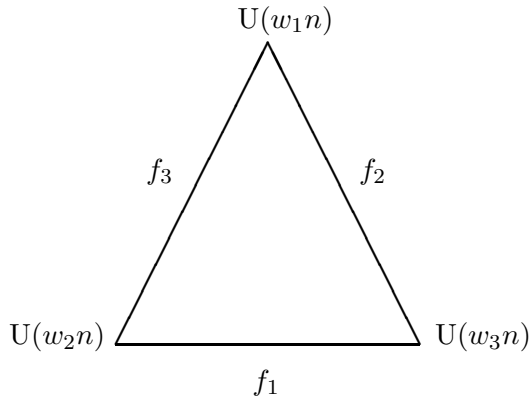
$$
 \mbox{
  \begin{picture}(100,132)(0,0)
  \unitlength=2cm
  \thicklines
   \put(0,0){\line(1,2){1}}
   \put(0,0){\line(1,0){2}}
   \put(2,0){\line(-1,2){1}}
   \put(0.8,2.1){$ \mbox{U}(w_1 n)$}
   \put(-0.7,0){$\mbox{U}(w_2 n)$}
   \put(2.1,0){$\mbox{U}(w_3 n)$}
   \put(0.9,-0.3){$f_1$}
   \put(1.6,1.1){$f_2$}
\put(0.2,1.1){$f_3$}
 \end{picture}
  }
  \label{four}
$$
The numbers $f_i$ are given by
\begin{equation}
\label{matter} f_i=\epsilon_{ijk}q^jq^kd
\end{equation}
where $d=\sum_i q_i$  which is  exactly the Calabi-Yau  condition in
${\bf WCP^2}_{q_1, q_2,q_3}$. \par In the end of this section, we
note  that it  is possible  to recover the result of GW from the
above toric description. As we have mentioned in the introduction,
the GW quiver has 3 vertices. Its vector of CS levels is $(k,0,-k)$.
Toric geometrically, this model  can be obtained by considering
$V^2$  as the canonical line bundle over $\bf CP^1$ which is  given
by
\begin{equation}
O(-2)\to \bf CP^1.
\end{equation}
The toric diagram for this geometry  is defined by  three vertices
$(v_0,v_1,v_2)$ such that
\begin{equation}
-2v_0+v_1+v_2=0.
\end{equation}
which is an open polytope.  This   can be represented  by 3 points
in $Z^2$  given by  ${v}_i=\{ (-1,1), (0,1), (1,1)\}$ where the last
entry is 1.  If we delete this entry we get  the mirror geometery
described by   ${v^\star}_i=\{(-1), (0), (1)\}$ along the line.
This toric data  can recover the GW quiver with three nodes. In this
case, the solution is given by
\begin{equation}
k_i={v^\star}_ik.
\end{equation}
More generally, this model  may be extended in a natural way to an
2d $\mathcal{N}=4$  $\mbox{U(1)}^r$ gauge theory  where  the moduli
space is given by the intersection of $r$  canonical line bundle
over $\bf CP^1$'s. The corresponding quiver may be identified with a
 bouquet of GW model.

\section{   More on toric   CS quivers }
The above  analysis may be extended to the case of  quivers with
more than three vertices. This can be  realized by replacing two
dimensional projective spaces  either by Hirzebrouch surfaces or del
Pezzo
 surfaces  whose toric diagrams  contain  more than three vertices.  The corresponding CS quivers involve
 more than three $\mbox{U}(N_i)$  gauge  symmetry  factors. Roughly,  Hirzebrouch surfaces ${\bf F}_\ell$,
 for instance,  are complex  two-dimensional toric
 surfaces  defined by non-trivial fibrations of
 a $\bf  CP^1$ over  a $ \bf CP^1$.  This fibration is  classified by
an  integer $\ell$, being the first Chern class integrated over $\bf
CP^1$. In $\mathcal{N}=2$ sigma model language, ${\bf F}_\ell$  are
realized as  vacuum manifolds  described by
         $ \mbox{U}(1) \times \mbox{U}(1)$ gauge theory with four chiral fields with charges
         \begin{eqnarray}
 q_i^{1}=(1,1,0,\ell), \qquad
q_i^{2}=(0,0,1,1).
\end{eqnarray}
Toric geometrically, these surfaces  are  represented by four
vertices $v_i$
          belonging to  $\bf Z^2$ and  satisfying the following toric constraints
 \begin{eqnarray}
\label{fn}
{ v_1}+{ v_2}+\ell{ v_4}=0\nonumber\\
{ v_3}+{ v_4}=0.
\end{eqnarray}
For simplicity reason,  we will restrict ourselves to
   ${\bf F}_0$   describing   a  trivial fibration of  $\bf  CP^1$ over   $ \bf CP^1$. In this case,
    eq.(\ref{fn}) reduces to
 \begin{eqnarray}
\label{f0}
{v_1}+{ v_2}=0\nonumber\\
{v_3}+{ v_4}=0.
\end{eqnarray}
Since we have more than one toric equation, the link  discussed  in
the previous section  is not obvious. However, the  corresponding
quiver can be interpreted as two orthogonal GW quiver models. In
terms of CS quivers, eq.(\ref{f0}) may be thought of as a particular
situation of
\begin{equation}
\label{conf0} k_1+k_2+k_3+k_4=0
\end{equation}
describing  a rectangular CS quiver with  the same gauge group ranks
\cite{MS}.  A priori,  this equation
  can  have may solutions. However,  toric geometry of two dimensional varieties  requires only  two possible
    solutions given by
\begin{eqnarray}
k_i=0,\qquad \sum\limits_{j\neq i}k_j=0,  \label{sol1} \\
k_i+k_j=0, \qquad \sum\limits_{k\neq i,j}k_k=0  \label{sol2}
\end{eqnarray}
where $i,j,k,l=1,2,3,4$. Otherwise, eq.(\ref{conf0}) can be
interpreted as a toric equation of three  dimensional
  projective space $\bf CP^3$ which is not needed  here.   It follows from the above discussion that the  first
   solution given in  eq.(\ref{sol1}) corresponds to  the CS quiver  based on  $\bf CP^2$ toric graph that corresponds to a GW model for two gauge factors with $(k,-k)$, whereas the second one
   eq.(\ref{sol2}) should  be associated with  CS quiver based  on  $F_0$ geometry.  The latter can be viewed as two orthogonal  GW models.
     Indeed, ignoring   the node associated  with  the  zero CS level, the corresponding CS quiver can be identified  with two blocks
       of two gauge factors with $(k,-k)$. The connection with $F_0$  requires a
   combination of the Mori vector  charges $q_i^a, (a=1,2)$.  The proposed  solution can be given by \begin{eqnarray}
k_i&=& (v^1_i+v^2_i)k\nonumber\\
N_i&=&(q^1_i+q^2_i)n.
\end{eqnarray}
It is possible to get the  CS quiver based on $F_0$ by  taking  the
following toric data
 \begin{eqnarray}
{ v_1}&=&(1,0)\nonumber\\
{ v_2}&=&(-1,0)\nonumber\\
{ v_3}&=&(0,1)\\
{ v_4}&=&(0,-1)\nonumber.
\end{eqnarray}
This toric data produces the quiver CS theory  proposed in
\cite{MS}. The gauge symmetry is given  by
\begin{equation}
\mbox{U} (N)\times\mbox{U}(N)\times \mbox{U}(N)\times \mbox{U}(N)
\end{equation}
where the CS levels are $(k,-k,k,-k)$. This model corresponds to a
circular quiver with four nodes, and it would be nice to see if this
quiver could be associated  with the quiver given in
\cite{imamura2}.
\section{ Discussions }
In this work, we  have discussed  3-dimensional $\mathcal{N}=4$ CS
quiver models. In particular, we have considered a class of such
models obtained by the compactification of  M-theory background
viewed as target space of $\mathcal{N}=4$ sigma model. The geometry
is realized explicitly as  cotangent bundles over  complex
  two-dimensional toric varieties. The corresponding quivers are identified with  toric graphs  of such two-Fano
   varieties.  In particular, we have analyzed in some
detail the cases of  two dimensional  projectives spaces and
Hirzebrouch surfaces. Using toric geometry,   we have discussed how
the physical constraints  of this  M-theory on backgrounds can be
related to toric description  of bi-dimensional  toric  manifolds.
We have discussed the possibility to  relate them with M2-brane
quivers models of the
 GW construction via restricting the CS constrains to those values that satisfy the Fundamental Identity as indicated in \cite{GW}.

\section*{Acknowledgments:}We are specially indebted to M. Asorey, J. L. Boya, F. Falceto 
 E. H. Saidi and particularly to D. Rodriguez-Gomez, for interesting discussion, clarifications and collaborations on related topics. This work is  partially
supported by   DGIID-DGA (grant 2011-E24/2),  CICYT (grant
FPA-2009-09638) and grant  A/93357/07, A/024147/09, A9335/10.  The
work of MPGM is funded by the Spanish Ministerio de Ciencia e
Innovaci\'on (FPA2006-09199) and the Consolider-Ingenio 2010
Programme CPAN (CSD2007-00042). AB would like to thank M. Asorey for
scientific help and AB and MPGM thanks to the Departamento de
F\'isica Te\'orica, Universidad de Zaragoza for kind hospitality. AB
would like to   thank also  L.J. Boya, Diaz Family, Montanez Family
and  Segui Family for kind hospitality in
Spain.

 \end{document}